\documentclass[twocolumn,preprintnumbers,amsmath,amssymb]{revtex4}
\usepackage{color}
\usepackage{graphicx}
\usepackage{dcolumn}
\usepackage{bm}
\begin{document}



\title{A Spinning Mirror for Fast Angular Scans of EBW Emission for 
Magnetic Pitch Profile Measurements\footnote{Contributed paper, published as part of the Proceedings of the 18th Topical Conference on High-Temperature Plasma Diagnostics, Wildwood, New Jersey, USA, May 2010.}}

\author{Francesco Volpe} \email{fvolpe@wisc.edu}
\affiliation{Department of Engineering Physics, University of Wisconsin, Madison, WI 53706, USA}

\date{\today}

\begin{abstract}
A tilted spinning mirror rapidly steers the line of sight of the 
electron Bernstein wave (EBW) emission radiometer at the Mega Amp Spherical 
Tokamak (MAST). 
In order to resist high mechanical stresses at rotation 
speeds of up to 12,000 rpm and to avoid eddy current induced magnetic braking, 
the mirror consists of a glass-reinforced nylon substrate of a special
self-balanced design, coated with a reflecting layer. 
By completing an angular scan every 2.5-10ms, it allows one 
to characterize with good time resolution the Bernstein-extraordinary-ordinary 
mode-conversion efficiency as a function of the view angles. 
Angular maps of conversion efficiency are directly related to the
magnetic pitch angle at the cutoff layer for the ordinary mode. Hence,
measurements at various frequencies provide the safety factor profile 
at the plasma edge. Initial measurements and indications of the feasibility 
of the diagnostic are presented. 
Moreover, angular scans indicate the best launch conditions for EBW heating. 
\end{abstract}

\maketitle

\section{INTRODUCTION and PHYSICAL PRINCIPLE} 
Measurements of the profile of safety factor $q$ in tokamaks are 
important for a number of stability studies 
(sawteeth, neo-classical tearing modes and resistive wall modes among others) 
as well as for internal transport barriers, non-inductive current drive and 
"advanced tokamak" and ``hybrid'' operation.

Means to measure the $q$ profile were reviewed in Ref.\cite{Rev1} and, more
recently, in Ref.\cite{Rev2}.
They include the Motional Stark Effect (MSE) \cite{MSE1, MSE2}, 
the Zeeman Effect in Lithium beam spectroscopy \cite{Zeeman2,Zeeman}, the 
heavy-ion-beam-probe \cite{HIBP} and arrayed line-integrated Faraday-rotation
measurements \cite{Faraday}.
Tangential Soft X-Ray imaging was also used to constrain equilibrium 
reconstructions and infer the $q$ profile \cite{SXR}.

In this article we present an alternative, compact and low-cost diagnostic 
of the edge $q$ profile based on radiometry of mode-converted 
electron Bernstein wave (EBW) emission. 
The diagnostic also identifies the best direction and conditions (edge
temperature, density gradient, etc.) for efficient EBW heating. 
The technique was conceived at the Mega Amp Spherical Tokamak (MAST) 
but doesn't rely on any spherical tokamak peculiarity and can thus 
in principle be exported to other overdense plasmas. 
Note also that it doesn't require the radiometer to be calibrated,  
neither absolutely nor relatively. 

EBWs have been excited by ordinary-extraordinary-Bernstein (OXB) 
mode conversion or detected via the reverse, BXO process, 
in several magnetic confinement devices \cite{LaquaRev}. 
The regions of existence for the O- and the slow
X-mode depend on the direction of propagation. A special direction minimizes 
the evanescent layer in between, thus favoring 
the tunneling from one region to the other. The
conversion efficiency $T$ degrades as the line of sight deviates from the
optimal direction \cite{Mjol}, as confirmed by angular scans of EBW heating
\cite{Laqua1}  and emission \cite{Laqua2} in one dimension. 
The conversion efficiency $T$, however, is $2(1+Y)$ 
times more sensitive to misalignment
perpendicular to the ambient magnetic field, than parallel to it
\cite{Mjol}. Here $Y$=$\omega_c/\omega$ is the dimensionless magnetic field and 
$\omega_c$ the electron cyclotron frequency. As a
consequence, iso-$T$ contours are non-circular, roughly elliptical
functions of the viewing angles.  
In tokamaks these ellipses are tilted, as a result of the field lines being 
tilted \cite{Cairns}. 
This was confirmed by two-dimensional scans performed at MAST: 
a multi-channel radiometer was connected to the EBW
heating antenna, but used in this case for emission measurements, and the line
of sight was steered manually in a number of reproducible discharges. $T$ was
observed to degrade with different rates as the line of sight deviated
toroidally (Fig.1a) or poloidally (Fig.1b) from optimal.

\begin{figure}[b]
   \includegraphics{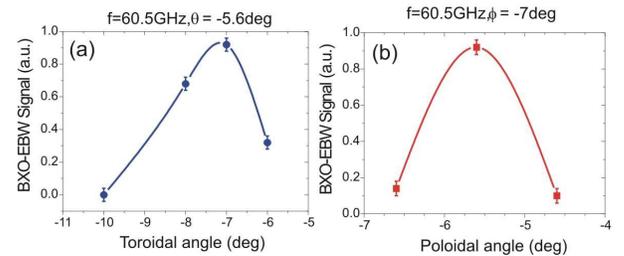} 
   \caption{Measurements of EBW emission undergoing B-X-O mode conversion, 
            for various (a) toroidal and (b) poloidal viewing angles 
            \cite{FST07}. 
            The full-width at half maximum amounts to 2.5 and 1.2$^{\rm o}$,
            respectively.}
\end{figure}

These scans also 
led to the identification of the conditions for strongest emission 
and, by reciprocity, for successful EBW heating of MAST \cite{FST07}. 
The results of Fig.1 motivated the development of a fast steering antenna in 
order to, respectively,
\begin{enumerate}
\item 
reconstruct the conversion efficiency contours as functions of the view angles 
and, from their shape and inclination, infer the
magnetic pitch angle at the OX conversion location, i.e. at the O-mode cutoff
layer for a certain frequency $f$. 
Simultaneous and/or scanning measurements at various f provide the pitch angle 
at various radii, i.e., ultimately, the $q$ profile, 
\item 
quickly assess the optimum conditions for heating within a single discharge. 
\end{enumerate}

The present paper presents the fast steering
antenna (Sec.2) and the first experimental results (Sec.3).

\begin{figure}[t]
   \includegraphics[width=80mm]{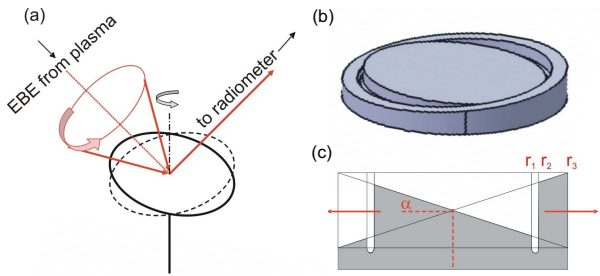}  
   \caption{    \label{FigSpMirr}
     (a) Principle of a tilted rotating mirror for angular scans. (b)
     Bird view and (c) cross-section of self-balanced mirror. Tilt angle 
   $a$ exaggerated for explanatory purposes.}
\end{figure}

\section{Self-balanced Spinning Mirror}                    
\begin{figure}[t]
   \includegraphics{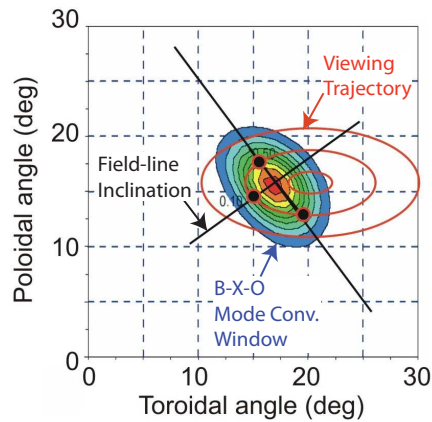}     
   \caption{\label{FigPrinciple80}
     BXO conversion efficiency contours for weak magnetic shear 
     (various colors) and angular 
   scan trajectories for mirrors tilted by $\alpha$=1.5, 3 and 4.5$^{\rm o}$ 
   (red), as functions of the view angles. Note that the conversion efficiency 
   contours are perpendicular to magnetic field lines.}
\end{figure}

Consider a rotating mirror tilted around the axis of rotation as in
Fig.\ref{FigSpMirr}a. 
This reflects into a radiometer horn a line-of-sight that changes with
time. Three mirrors were developed, with tilt angles $\theta$=1.5, 3 and 
4.5$^{\rm o}$, capable of horizontal scans of $\pm 2\theta$ 
and vertical scans of $\pm \theta$. Correspondingly,
the line of sight describes the elliptical trajectories marked in red in the
angular space of Fig.\ref{FigPrinciple80}. 
Before a discharge, pre-alignment of the diagnostic
sets the center of the angular scan; selecting a mirror of one tilt or another
determines the width of the scan. The choice depends on the type of discharge
(e.g. L-mode or H-mode) and the corresponding expected conversion efficiency
contours (narrow or broad, respectively). Good pre-settings allow to sample
the conversion efficiency contours during (a fraction of) the angular scan. In
particular it is desirable to cross in two points at least one of the axes of
symmetry of these elliptical contours, so that their inclination can be
determined. 

 The simplistic, unbalanced setup of Fig.\ref{FigSpMirr}a would be
subject to forces trying to flatten the mirror and annihilate the
tilt. To compensate for these forces, a rotor is manufactured as a
single block featuring a central reflector surrounded by a
counter-balancing ring of equal and opposite inclination, as in
Figs.\ref{FigSpMirr}b-c. The ring thickness, $r_3-r_2$, 
is related to the central
reflector radius $r_1$ by the simple relation $r_1^4= r_3^4- r_2^4$ so that
centrifugal forces of 3.3585 tonnes acting on one half of the rotor at
12,000rpm are compensated by equal and opposite forces acting on the
other half.

\begin{figure}[b]
   \includegraphics[scale=0.65]{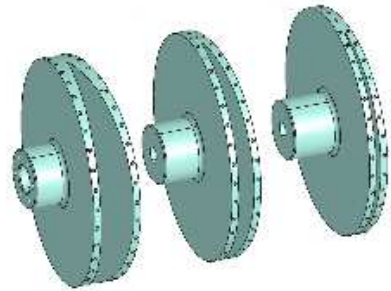}     
   \caption{\label{Figiso2}
     Alternative spinning mirror design tilted (from left to right) by 4.5, 
     3 and 1.5$^{\rm o}$.}
\end{figure}

In an alternative design (Fig.\ref{Figiso2}) the counter-balancing element 
lies behind the actual mirror, not around it. This, however, would be 
subject to higher orders of vibration. Hence, the design of 
Fig.\ref{FigSpMirr}c was adopted. 

The material adopted, glass-fiber-reinforced Nylon 66, avoids magnetic
braking and other eddy current effects and withstands the high
mechanical stresses involved. It also has relatively low density,
which reduces the centrifugal forces. The diagnostic support was
carefully designed and modeled to avoid mechanical resonances at the
frequencies of interest (100-200Hz). Vibrations, already reduced by
the special design, were further minimized with the help of a
balancing machine. As a result, excellent balancing was demonstrated,
with forces as little as 1N even at full speed (12,000rpm) and for the
steepest (4.5$^{\rm o}$), thus, potentially most unbalanced spinning
mirror. Note that, due to centrifugal forces, the shape of the rotor
changes with speed. As a result, optimal balancing for a certain speed
is not necessarily optimal at other velocities, including lower
speeds.

The Nylon substrate was successfully polished (despite the
difficulties associated with the glass fibers embedded in it) {\em or}
coated with gold. However, coating on the polished material tended to
peel off. For this reason, a layer of gold as thick as 3 skin depths
was deposited on a 2mm polished polycarbonate plate and this was glued
with strong epoxy on the Nylon, which was re-balanced. Each mirror was
mounted on a shaft held by a matched pair of ball bearings and by a
cylindrical roller bearing in a spindle housing. Three such 
``units'' were prepared, with mirrors inclined by 1.5, 3 and 4.5$^{\rm o}$. A
3-phase induction motor drives the shaft. The motor and the shaft are not 
coupled rigidly, but by pulleys and a flat belt. Units can be substituted
before a plasma discharge, depending on the desired angular scan
width. The orientation of the mirror is measured with 1$^{\rm o}$ precision by
an optical encoder.

\begin{figure}[t]
   \includegraphics{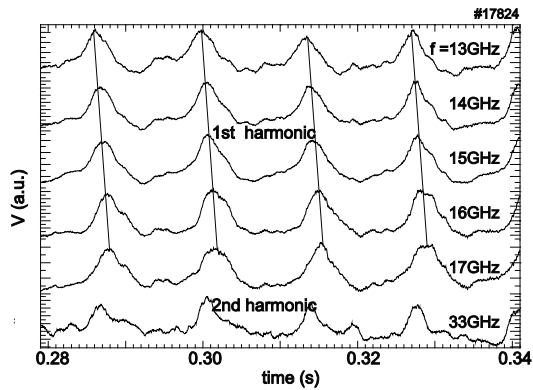}     
   \caption{\label{FigExpRes1}
     EBW emission at various frequencies, during rapid (13ms) 2D
     angular scans. Different channels reach the maximum (i.e., maximum
     mode conversion) at different times (i.e. for different angles), in
     agreement with the different magnetic pitch angle at the cutoff.}
\end{figure}

\section{Experimental Results} 
The manual, shot-by-shot angular scans of Fig.1 were performed with a
12 channel, 54-66GHz radiometer connected to the 60GHz EBW heating
antenna, whose steering mirrors are on the vacuum side.  

The new fast steering antenna, however, has been installed outside the torus, 
in front of the main EBW emission diagnostic. 

The spinning mirror 
reflects the EBW emission exiting from the torus flange onto a flat
mirror and this in a horn antenna and a Teflon lens optimized for
40GHz that refocuses the beam and limits its divergence to less than
3$^{\rm o}$. All this Gaussian optics is located 20cm below the plasma
mid-plane. Directional couplers split the signal between three 
scanning radiometers (16-26, 26-40 and 40-60GHz), here operated at fixed 
frequencies, and a bank of 11 fixed channels (6-18GHz). 
The measurements presented here were carried out with the 
4.5$^{\rm o}$ spinning mirror in the 10-36 GHz range. 
These frequencies correspond to {\em emission} at the first and second harmonic 
and to OX {\em conversion} at locations where 
$n_e=1.2\cdot 10^{18}$-$1.6\cdot 10^{19}m^{-3}$, i.e.~at the plasma edge. 

Fig.\ref{FigExpRes1} 
shows the first angular scans of EBW emission realized with the spinning
mirror within a single discharge. Although optimized for 12,000rpm,
for this initial test the mirror was rotated at 4400rpm, yielding a
complete angular scan for each channel every 13.6ms. As expected,
different channels reach their maximum  at different times, i.e. for
different view angles. This is consistent with the emission at
different frequencies undergoing the final XO conversion at distinct
cutoff layers, lying at different depths in the plasma, where they
pick up the information on the local pitch angle. 
As the local pitch angle changes from channel to channel, the optimal 
direction also changes, and the best alignment is obtained at different times. 
The trend is consistent for all channels looking at the 1st harmonic, then 
the time-delay is ``reset'' in the transition to the 2nd harmonic, consistently 
with the fact that the emission is suddenly originating in an outer position. 

\begin{figure}[t]
   \includegraphics{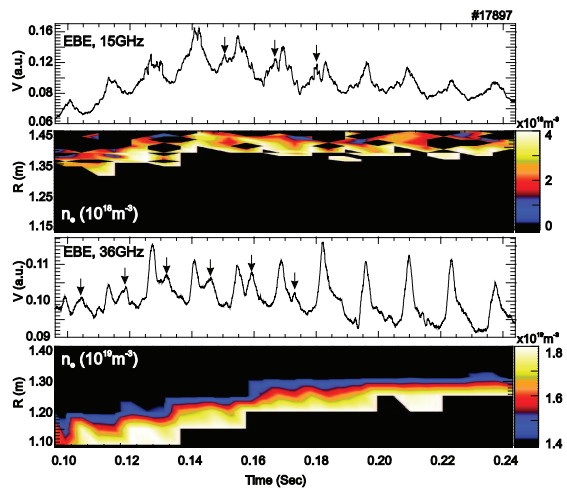}     
   \caption{\label{FigExpRes2}
     Evidence of a second peak (arrowed) for each angular scan
     of EBW emission at the 1st (a) and 2nd (c) cyclotron
     harmonic. (b, d) Density contours suggest that the density scale-length 
     remains constant, 
     but the cutoff layer for 15GHz is turbulent and the cutoff for 36GHz 
     drifts in time before reaching a steady position, possibly explaining
     intermittent observation of the second peak at 15GHz and no second
     peak after t=0.18s at 36GHz.}
\end{figure}

Additional effects contribute to the time-modulation of the signal and need 
to be minimized or deconvolved:  
\begin{itemize}
\item
incomplete shine-through of the
microwave beam, slightly broader and thus partly masked by the port,
especially at low frequencies, 
\item
stray radiation reaching the antenna 
after multiple reflections in the vessel, 
\item
changes of the Doppler
shift with the view angle. As a result, the EBW emitting layers move
radially by a few centimeters during the scan. The resulting
oscillation of the radiative temperature, however, amounts to a few
percent, and is dwarfed by the change in conversion efficiency, which
can oscillate by as much as 100\%.  
\end{itemize}

However, these effects alone would cause all channels to oscillate with the 
same phase. Time-shifts from channel to channel, as observed in 
Fig.\ref{FigExpRes1}, are considered a signature of genuine angular scan of 
BXO-converted EBW emission at different frequencies.

\begin{figure}[t]
   \includegraphics{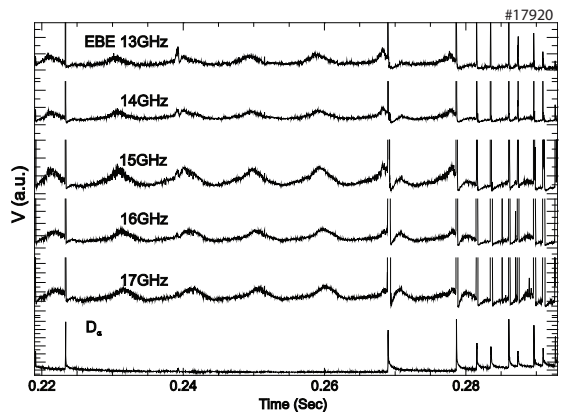}     
   \caption{\label{FigExpRes3}
     EBW emission correlates with ELMs. A complete angular scan
     is only possible if the inter-ELM period is sufficiently long.}
\end{figure}

The data in question oscillate at the rotation frequency of the
mirror, suggesting that they cross the conversion efficiency contours once 
per revolution. Hints of a second peak are visible in Fig.\ref{FigExpRes1} but
become more evident in another discharge, in Fig.\ref{FigExpRes2}, with a 
steeper density gradient and a broad enough conversion efficiency window for
the angular scan to cross its axis twice. Time intervals with two
maxima are good candidates for inferring the inclination of the
conversion efficiency contours and thus, ultimately, the magnetic
pitch angle. The analysis is in progress.  The mode-converted emission
appears to correlate with edge localized modes (ELMs) (Fig.\ref{FigExpRes3}). 
Like Electron Cyclotron Emission, pure EBW
emission is expected indeed to be correlated with ELMs. The mode
conversion, instead, becomes less efficient during an ELM, due to the
reduced density gradient. In other words, ELMs enhance 
EBW emission but reduce BXO transmission.  
The signals are the product of these two quantities. 
Note that emission doesn't depend strongly on the view angles, 
while transmission does.
%
Note also that a pitch angle measurement needs a complete
angular scan, and this is only possible if the inter-ELM period is
sufficiently long, as for $t$=0.225-0.265s in Fig.\ref{FigExpRes3}. 

\vspace*{5mm}
\section{Summary and Conclusions}

In conclusion a technique for measuring the $q$ profile was 
proposed, based on anisotropy of the BXO mode conversion 
and a tilted spinning mirror of special self-balanced design 
was installed on MAST for this purpose: 
by completing an angular scan of EBW emission every 2.5-10ms, the spinning 
mirror allows to characterize with good time resolution the BXO 
conversion efficiency as a function of the view angles. 
This decays anisotropically as the line of sight departs from an optimal 
direction. In first approximation the direction of steepest descent gives the 
local magnetic field. In presence of magnetic shear, however, the contours 
deform \cite{Cairns} and a more 
sophisticated analysis involving a full-wave code would become necessary. 
The device was described, with emphasis on the materials and design. The 
first angular scan results exhibit the expected shape, 
frequency dependence and ELM dependence.

\begin{acknowledgments}
This work was funded jointly by the United Kingdom Engineering and Physical
Sciences Research Council and by the European Communities under the contract
of Association between EURATOM and CCFE. The views and opinions expressed
herein do not necessarily reflect those of the European Commission.
\end{acknowledgments}



\end{document}